\documentclass[fleqn,12pt,twoside]{article}
\usepackage[headings]{espcrc1}

\readRCS
$Id: espcrc1.tex,v 1.2 2004/02/24 11:22:11 spepping Exp $
\usepackage{graphicx}
\usepackage[figuresright]{rotating}
\usepackage{color}

\newcommand{\AmS}{{\protect\the\textfont2
  A\kern-.1667em\lower.5ex\hbox{M}\kern-.125emS}}

\title{Results from NA49}

\author{C.~H\"ohne (for the NA49 Collaboration) \\
\vspace{0.3cm}
C.~Alt$^{9}$, T.~Anticic$^{21}$, B.~Baatar$^{8}$,D.~Barna$^{4}$,
J.~Bartke$^{6}$, L.~Betev$^{10}$, H.~Bia{\l}\-kowska$^{19}$,
C.~Blume$^{9}$,  B.~Boimska$^{19}$, M.~Botje$^{1}$,
J.~Bracinik$^{3}$, R.~Bramm$^{9}$, P.~Bun\v{c}i\'{c}$^{10}$,
V.~Cerny$^{3}$, P.~Christakoglou$^{2}$, O.~Chvala$^{14}$,
J.G.~Cramer$^{16}$, P.~Csat\'{o}$^{4}$, P.~Dinkelaker$^{9}$,
V.~Eckardt$^{13}$,
D.~Flierl$^{9}$, Z.~Fodor$^{4}$, P.~Foka$^{7}$, V.~Friese$^{7}$,
J.~G\'{a}l$^{4}$, M.~Ga\'zdzicki$^{9,11}$, V.~Genchev$^{18}$,
G.~Georgopoulos$^{2}$, E.~G{\l}adysz$^{6}$, K.~Grebieszkow$^{20}$,
S.~Hegyi$^{4}$, C.~H\"{o}hne$^{7}$, K.~Kadija$^{21}$,
A.~Karev$^{13}$, M.~Kliemant$^{9}$, S.~Kniege$^{9}$,
V.I.~Kolesnikov$^{8}$, E.~Kornas$^{6}$, R.~Korus$^{11}$,
M.~Kowalski$^{6}$, I.~Kraus$^{7}$, M.~Kreps$^{3}$,
A.~Laszlo$^{4}$, M.~van~Leeuwen$^{1}$, P.~L\'{e}vai$^{4}$,
L.~Litov$^{17}$, B.~Lungwitz$^{9}$, M.~Makariev$^{17}$,
A.I.~Malakhov$^{8}$, M.~Mateev$^{17}$, G.L.~Melkumov$^{8}$,
A.~Mischke$^{1}$, M.~Mitrovski$^{9}$, J.~Moln\'{a}r$^{4}$,
St.~Mr\'owczy\'nski$^{11}$, V.~Nicolic$^{21}$, G.~P\'{a}lla$^{4}$,
A.D.~Panagiotou$^{2}$, D.~Panayotov$^{17}$, A.~Petridis$^{2}$,
M.~Pikna$^{3}$, D.~Prindle$^{16}$, F.~P\"{u}hlhofer$^{12}$,
R.~Renfordt$^{9}$, C.~Roland$^{5}$, G.~Roland$^{5}$, M.
Rybczy\'nski$^{11}$, A.~Rybicki$^{6,10}$, A.~Sandoval$^{7}$,
N.~Schmitz$^{13}$, T.~Schuster$^{9}$, P.~Seyboth$^{13}$,
F.~Sikl\'{e}r$^{4}$, B.~Sitar$^{3}$, E.~Skrzypczak$^{20}$,
G.~Stefanek$^{11}$, R.~Stock$^{9}$, C.~Strabel$^{9}$,
H.~Str\"{o}bele$^{9}$, T.~Susa$^{21}$, I.~Szentp\'{e}tery$^{4}$,
J.~Sziklai$^{4}$, P.~Szymanski$^{10,19}$, V.~Trubnikov$^{19}$,
D.~Varga$^{4,10}$, M.~Vassiliou$^{2}$, G.I.~Veres$^{4,5}$,
G.~Vesztergombi$^{4}$,
D.~Vrani\'{c}$^{7}$, A.~Wetzler$^{9}$, Z.~W{\l}odarczyk$^{11}$
I.K.~Yoo$^{15}$, J.~Zim\'{a}nyi$^{4}$\\
\vspace{0.3cm}
$^{1}$NIKHEF, Amsterdam, Netherlands. \\
$^{2}$Department of Physics, University of Athens, Athens, Greece.\\
$^{3}$Comenius University, Bratislava, Slovakia.\\
$^{4}$KFKI Research Institute for Particle and Nuclear Physics, Budapest, Hungary.\\
$^{5}$MIT, Cambridge, USA.\\
$^{6}$Institute of Nuclear Physics, Cracow, Poland.\\
$^{7}$Gesellschaft f\"{u}r Schwerionenforschung (GSI), Darmstadt, Germany.\\
$^{8}$Joint Institute for Nuclear Research, Dubna, Russia.\\
$^{9}$Fachbereich Physik der Universit\"{a}t, Frankfurt, Germany.\\
$^{10}$CERN, Geneva, Switzerland.\\
$^{11}$Institute of Physics \'Swi{\,e}tokrzyska Academy, Kielce, Poland.\\
$^{12}$Fachbereich Physik der Universit\"{a}t, Marburg, Germany.\\
$^{13}$Max-Planck-Institut f\"{u}r Physik, Munich, Germany.\\
$^{14}$Institute of Particle and Nuclear Physics, Charles University, Prague, Czech Republic.\\
$^{15}$Department of Physics, Pusan National University, Pusan, Republic of Korea.\\
$^{16}$Nuclear Physics Laboratory, University of Washington, Seattle, WA, USA.\\
$^{17}$Atomic Physics Department, Sofia University St. Kliment Ohridski, Sofia, Bulgaria.\\
$^{18}$Institute for Nuclear Research and Nuclear Energy, Sofia, Bulgaria.\\
$^{19}$Institute for Nuclear Studies, Warsaw, Poland.\\
$^{20}$Institute for Experimental Physics, University of Warsaw, Warsaw, Poland.\\
$^{21}$Rudjer Boskovic Institute, Zagreb, Croatia.}

\runtitle{Results from NA49} \runauthor{C.~H\"ohne (for the NA49
Collaboration)}

\begin{document}

\maketitle

\begin{abstract}
An overview of results from the CERN experiment NA49 is presented
with emphasis on most recent measurements. NA49 has systematically
studied the dependence of hadron production on energy and system
size or centrality. At top-SPS energy the detailed investigation of
hadron production, now also extending to elliptic flow of
$\Lambda$-baryons and to identified particle yields at high $p_{t}$,
shows that the created matter behaves in a similar manner as at RHIC
energies. In the lower SPS energy range a distinct structure is
observed in the energy dependence of the rate of strangeness
production and in the slopes of $p_{t}$-spectra suggesting the onset
of the creation of a deconfined phase of matter.

\end{abstract}

\section{Introduction}

Over the 9 years of data taking at the CERN-SPS (1994-2002), NA49
has collected a comprehensive set of data on hadron production in
A+A collisions aiming at an understanding of the reaction mechanism
through a systematic study of many observables in different systems
and at different energies. Distinct changes of hadron production
properties are expected once a deconfined phase of matter is created
in the early stage of the collision.

Main observables in addition to a large variety of particle
correlation and fluctuation measures were the hadro-chemical
composition of the final state, in particular the strangeness
content. The most extensive studies were performed for central Pb+Pb
collisions at the top-SPS beam energy of 158$A$ GeV, partially also
extending to larger impact parameters and smaller collision systems.
For many observables the energy dependence was studied over the full
SPS energy range now providing a nearly continuous coverage from
threshold to RHIC energies. This energy scan revealed that the SPS
energy range is a particularly interesting region showing signs for
the onset of deconfinement.

The NA49 detector \cite{na49_nim} is a large acceptance hadron
spectrometer. Main components are four large volume time projection
chambers two of them being located inside a magnetic field. Particle
identification is done by a measurement of the specific energy loss
in the TPCs, the time-of-flight around midrapidity, and by the study
of decay topology and invariant mass. The centrality of the
collisions is determined from the spectator energy measured in a
forward calorimeter downstream of the spectrometer. The
corresponding number of wounded nucleons $N_{\rm{wound}}$ and
participating nucleons $N_{\rm part}$ is extracted by simulations
based on the VENUS model \cite{venus}. In contrast to the
calculation of $N_{\rm{wound}}$, nucleons participating through
secondary cascading processes are also included in $N_{\rm part}$.

\section{Hadron production at 158$A$ GeV beam energy}

In central Pb+Pb collisions at 158$A$ GeV beam energy matter of high
energy density, certainly exceeding 1 GeV/fm$^{3}$, is created at
the early stage of the collision \cite{na49_e-density}. The yields
of produced hadrons can successfully be described by hadron gas
models, see e.g.\,\cite{becattini,pbm}, requiring temperatures of
the order of 160 MeV, a baryochemical potential of about 240 MeV
and, depending on the model, a strangeness undersaturation factor
$\gamma_{s}$. These conditions show that the hadrochemical
freeze-out occurs close to the predicted phase boundary
\cite{lattice} as is also the case for collisions at the higher RHIC
energies \cite{rhic_freeze-out}. It can be argued that already this
is an indication that a deconfined state is created prior to
freeze-out \cite{stock,wetterich}.

\begin{figure}[htb]
\begin{minipage}[t]{75mm}
\includegraphics[width=75mm]{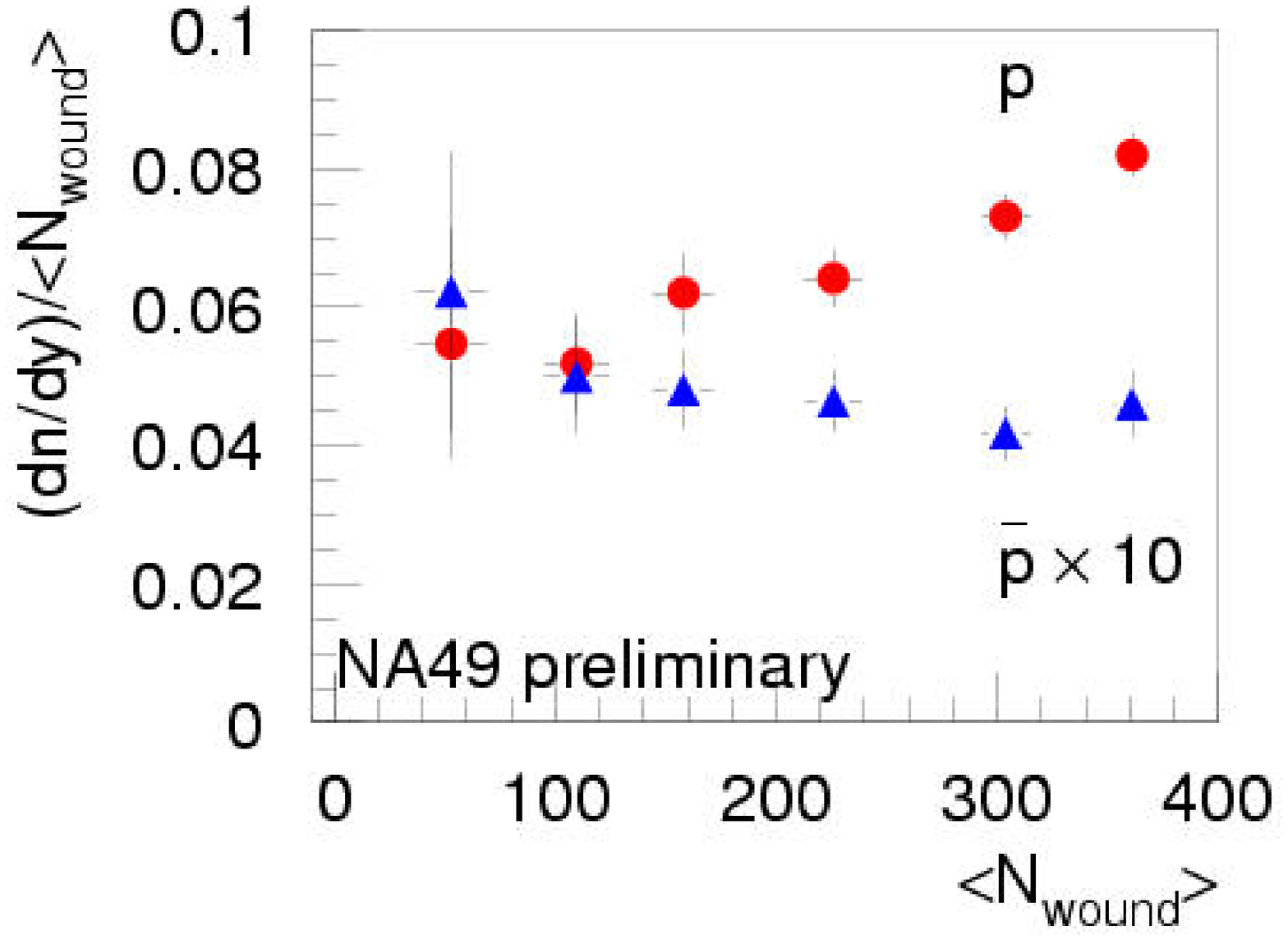}
\vspace{-1.4cm} \caption{Proton and antiproton yields at midrapidity
($2.4<y<2.8$) normalized to the number of wounded nucleons
$N_{\rm{wound}}$ versus centrality for Pb+Pb collisions at 158$A$
GeV.} \label{fig:ppbar_centrality}
\end{minipage}
\hspace{\fill}
\begin{minipage}[t]{75mm}
\includegraphics[width=75mm]{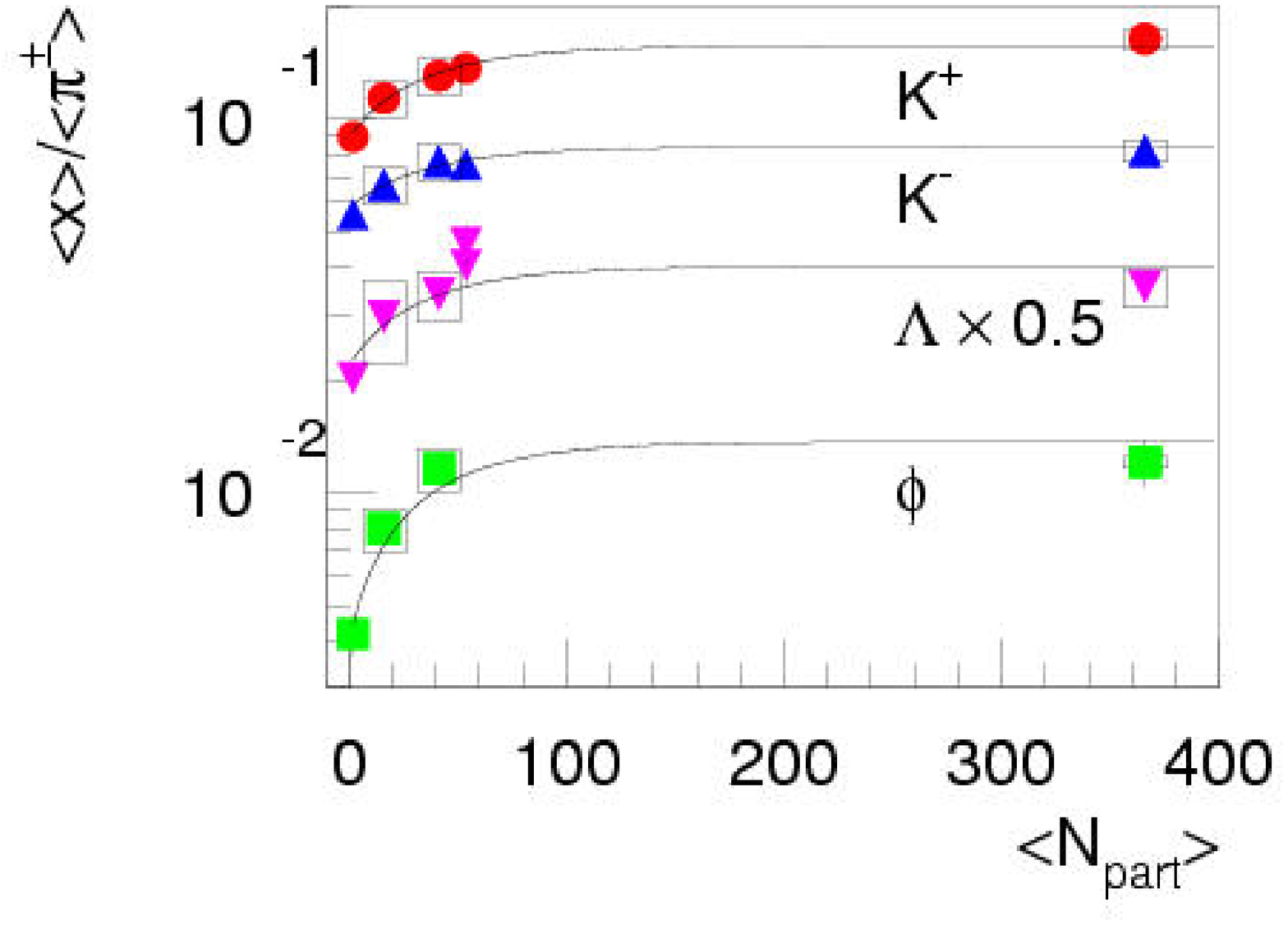}
\vspace{-1.4cm}\caption{Strange hadron yields in $4\pi$ normalized
to pions ($\langle\pi^{\pm}\rangle=\frac{\langle\pi^{+}\rangle
+\langle\pi^{-}\rangle}{2}$) for increasingly large collision
systems (p+p, central C+C, Si+Si, S+S and Pb+Pb)
\cite{na49_system-size}; lines are to guide the eye.}
\label{fig:s_system-size}
\end{minipage}
\end{figure}

An effect of the higher baryon densities reached at top-SPS energies
compared to RHIC is e.g.\,seen in the centrality dependence of
midrapidity proton and antiproton yields normalized to the number of
wounded nucleons (fig.~\ref{fig:ppbar_centrality}). A slow increase
is observed for protons which can be understood as a consequence of
the increased stopping in central collisions. On the other hand,
antiproton yields show a slight decrease which could be a sign of
increasing antiproton absorption.

In contrast, a fast increase with system size is observed for
relative strangeness production (see fig.~\ref{fig:s_system-size})
reaching a saturation at about 60 participants
\cite{na49_system-size}; similar results are also observed at RHIC
for the centrality dependence of Au+Au collisions
\cite{phenix_s_centrality}. Although this phenomenon is
qualitatively expected from statistical models due to strangeness
suppression in small volumes, it cannot be explained quantitatively
assuming a simple proportionality between volume and number of
participants \cite{redlich}. However, if one assumes several
coherence volumes in the small collision systems as suggested by
percolation calculations, the rise over a larger range of $N_{\rm
part}$ in the data can be described also quantitatively
\cite{hoehne}.

\begin{figure}[htb]
\begin{minipage}[t]{75mm}
\includegraphics[width=75mm]{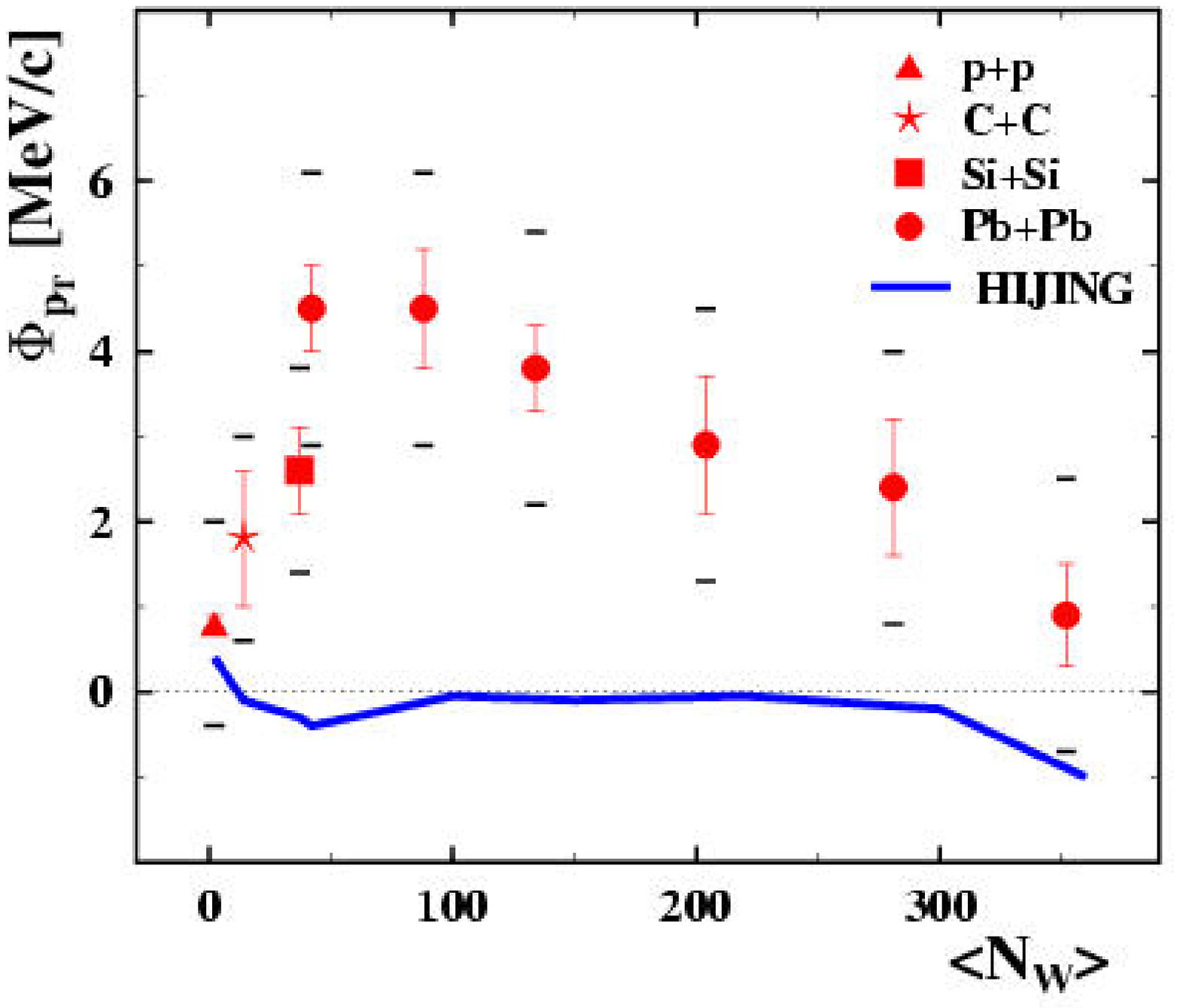}
\vspace{-1.4cm}\caption{$\phi_{p_{t}}$ measure of $\langle
p_{t}\rangle$ fluctuations for negatively charged particles versus
system size (p+p, central C+C, Si+Si, Pb+Pb) and centrality (Pb+Pb)
at 158$A$ GeV ($4<y_{\pi}<5.5$) \cite{na49_mean-pt}.}
\label{fig:mean-pt_fluctuations}
\end{minipage}
\hspace{\fill}
\begin{minipage}[t]{75mm}
\includegraphics[width=75mm]{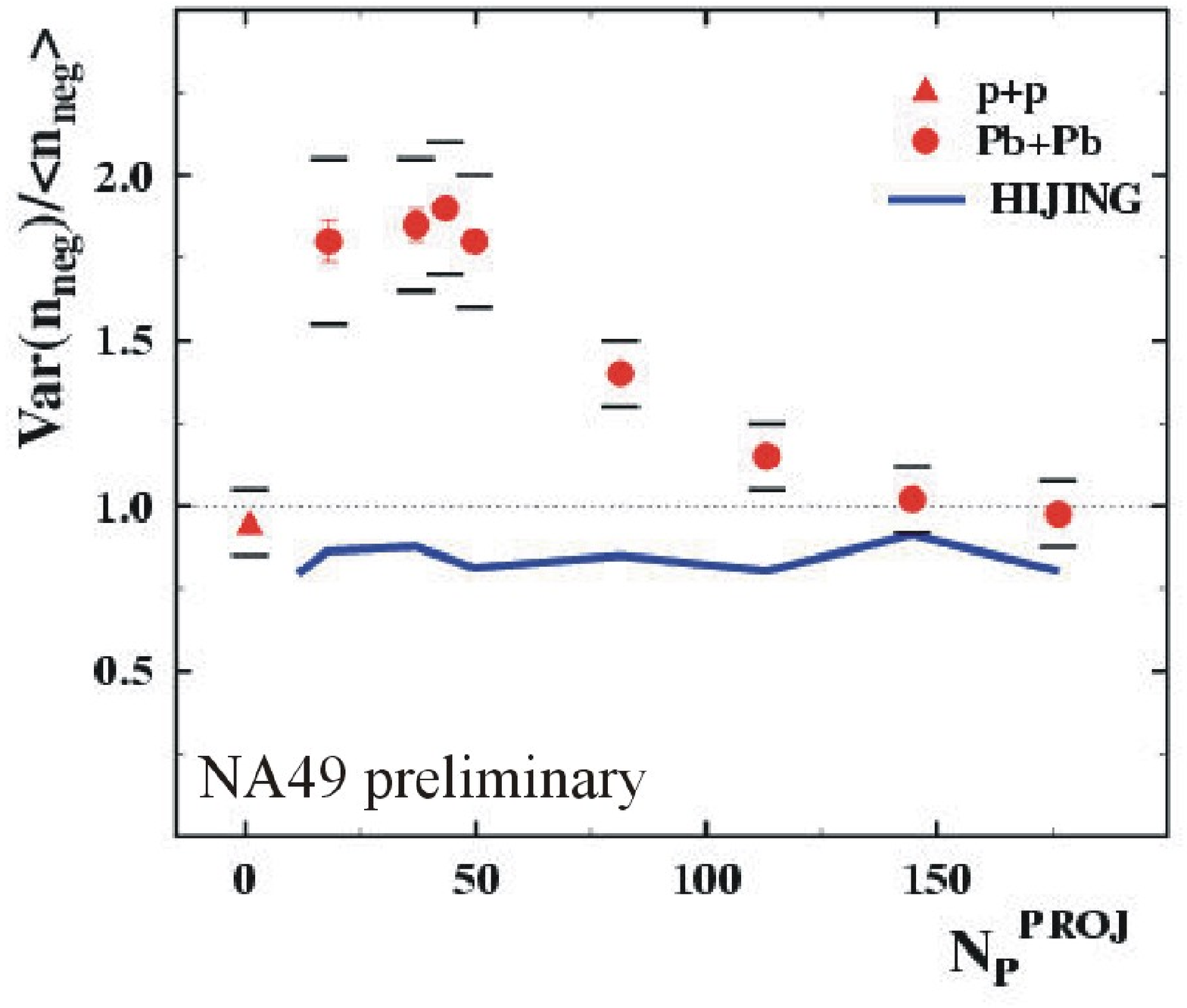}
\vspace{-1.4cm}\caption{Multiplicity fluctuations for negatively
charged particles versus centrality for Pb+Pb collisions at 158$A$
GeV beam energy ($4<y_{\pi}<5.5$).} \label{fig:mult_fluctuations}
\end{minipage}
\end{figure}

Within such a percolation picture the observed increase of $\langle
p_{t}\rangle$ and multiplicity fluctuations
(fig.~\ref{fig:mean-pt_fluctuations} and
\ref{fig:mult_fluctuations}) towards smaller collision systems can
also be understood \cite{ferreiro}. Dynamical fluctuations would
increase in the presence of several differently sized clusters but
would be small if essentially a single string (p+p) or one large
cluster (central Pb+Pb) existed. Assuming the known correlation of
$\langle p_{t} \rangle$ and charged multiplicity in p+p reactions
also for A+A collisions, the observed $\langle p_{t} \rangle$
fluctuations can be derived from the measured multiplicity
fluctuations \cite{mrowczynski}. A similar centrality dependence of
$\langle p_{t}\rangle$ fluctuations was observed at RHIC energies
\cite{phenix_mean-pt}.

\begin{figure}[htb]
\begin{minipage}[t]{75mm}
\includegraphics[width=75mm]{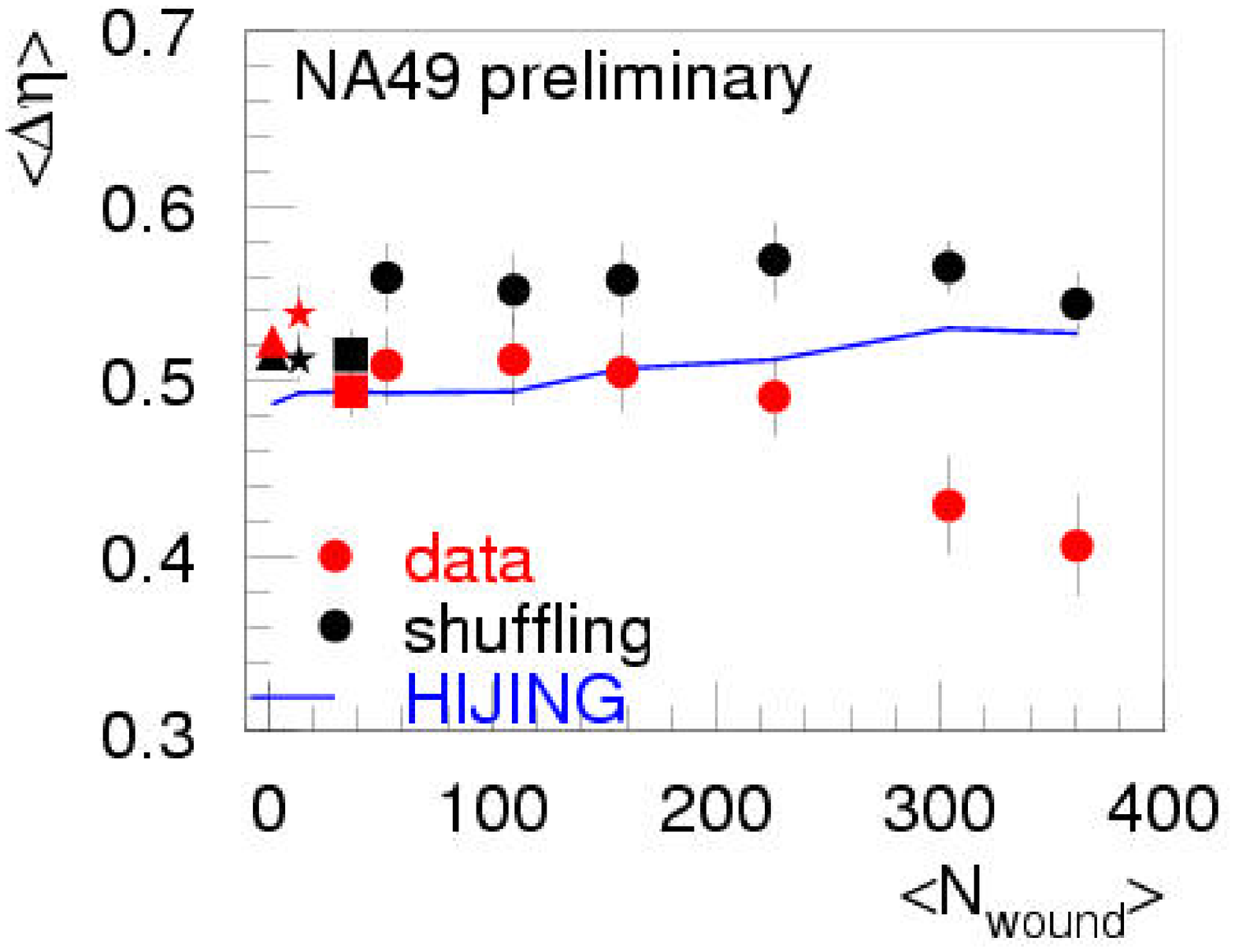}
\end{minipage}
\hspace{\fill}
\begin{minipage}[t]{75mm}
\includegraphics[width=75mm]{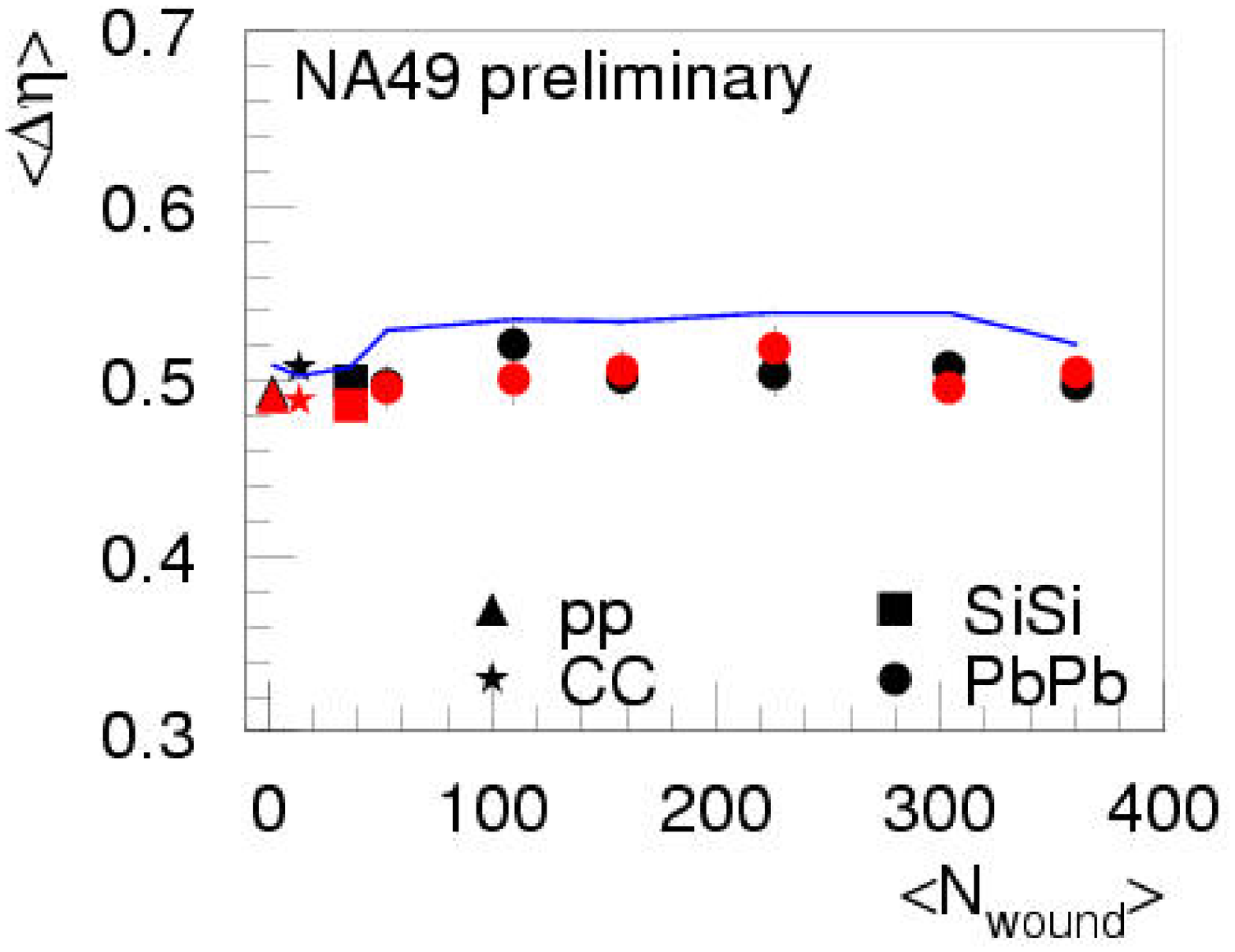}
\end{minipage}
\vspace{-1cm}\caption{Width $\langle\Delta\eta\rangle$ of the
balance function versus system size and centrality for charged
particles at mid-pseudorapidity $2.5 < \eta < 3.9$ (left) and for
forward pseudorapidities $4<\eta < 5.4$ (right) in p+p and A+A
collisions at 158$A$ GeV.} \label{fig:bf}
\end{figure}

Additional information on the reaction mechanism can be extracted
from the analysis of particle correlations, e.g.\,via the charge
balance function. Similar to what has been measured at RHIC
\cite{star_bf}, NA49 observes a decrease of the width of the balance
function for charged particles towards central Pb+Pb collisions.
This is consistent with a delayed hadronization scenario in central
Pb+Pb compared to collisions of larger impact parameter or of
smaller systems \cite{na49_bf}. Such a scenario is also supported by
an analysis of chemical and kinetic freeze-out conditions in the
smaller systems \cite{ingrid_taos}. Separate measurements of the
width at midrapidity and at forward rapidities (fig.~\ref{fig:bf})
reveal that the narrowing effect is located at midrapidity.

Recently, NA49 also started to investigate observables which so far
have been the domain of RHIC as e.g.\,elliptic flow of strange
particles or high-$p_t$ phenomena. Remarkably, the similarities
between the matter created at top-SPS and RHIC also extend to these
observations: In Pb+Pb collisions at 158$A$ GeV a substantial
elliptic flow for $\Lambda$-baryons is found which is similar to
that of protons but clearly smaller than that of pions
\cite{na49_lambda-flow}. The measurement of particle yields at high
$p_{t}$ shows a strong increase of the baryon/meson ratio at $p_t >
2.5$~GeV/c \cite{na49_high-pt}.

\section{Energy dependence of hadron production}

Motivated by the hypothesis \cite{marek} that the phase transition
is first crossed between AGS and top-SPS energies, NA49 has measured
hadron production in 5 energy steps from 20 to 158$A$ GeV. A hadron
gas analysis of the produced large set of hadron yields at each
energy shows that towards lower energies the freeze-out temperature
decreases while the baryochemical potential increases
\cite{becattini}. Comparing this smooth change of parameters with
the phase boundary calculated from lattice QCD \cite{lattice} it
appears that for lower energies the systems at freeze-out start
departing from this boundary.
\begin{figure}[htb]
\begin{minipage}[t]{75mm}
\includegraphics[width=53mm]{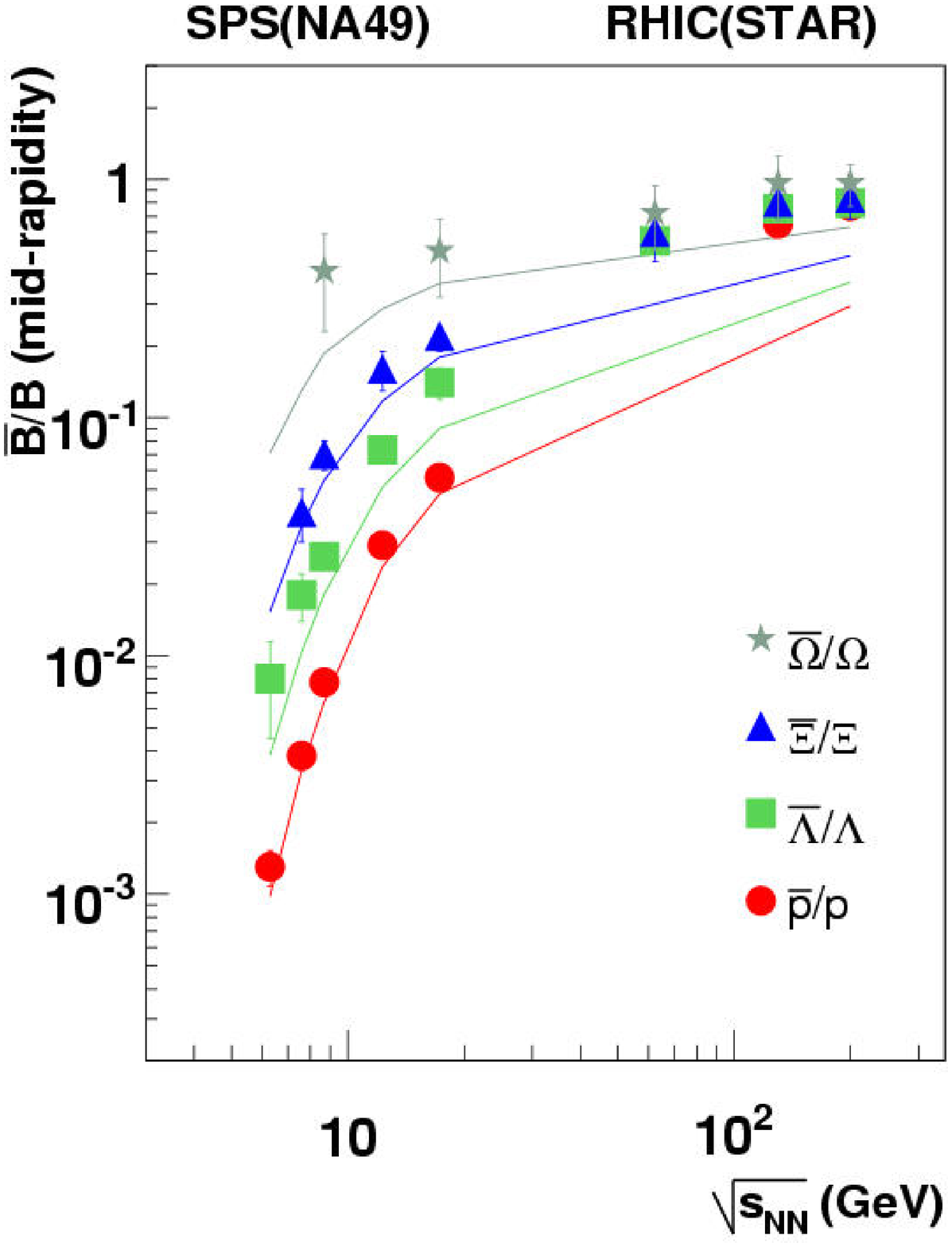}
\vspace{-1cm}\caption{$\overline{\rm{B}}$/B ratio at midrapidity
versus energy. The lines show the ratio from a hadron gas model fit
to $4\pi$ data including a strangeness undersaturation factor
$\gamma_{s}$ \cite{manninen}.} \label{fig:bbar-ratios}
\end{minipage}
\hspace{\fill}
%
\begin{minipage}[t]{75mm}
\includegraphics[width=58mm]{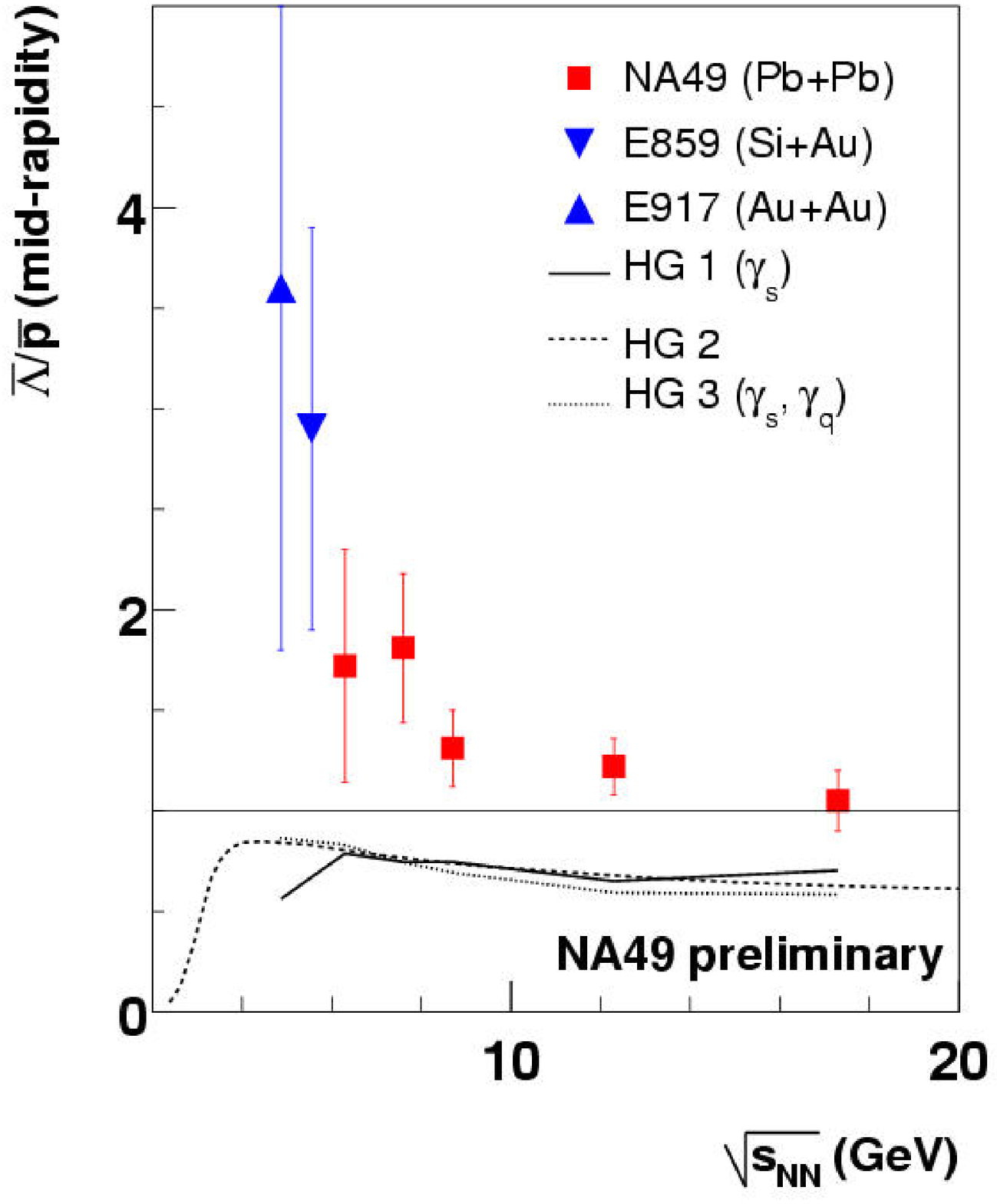}
\vspace{-1cm}\caption{$\overline{\Lambda}/\overline{\rm {p}}$
ratio at midrapidity versus energy ($\overline{\Lambda}$ not
corrected for contributions from $\overline{\Xi}$ decay). The
lines show the ratio from hadron gas model fits to $4\pi$ data: HG
1 including $\gamma_{s}$ \cite{manninen}, HG 2 without
\cite{hg-redlich}, HG 3 including $\gamma_{s}$ and $\gamma_{q}$
\cite{rafelski}.} \label{fig:alambda-ap-ratio}
\end{minipage}
\end{figure}

\begin{figure}[htb!]
\begin{minipage}[t]{115mm}
\includegraphics[width=110mm]{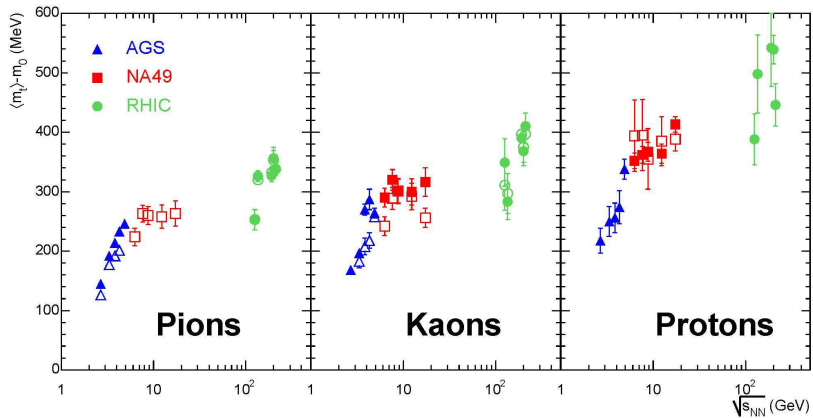}
\end{minipage}
\hspace{\fill}
\begin{minipage}[t]{35mm} \vspace{-5.3cm}
\caption{Energy dependence of $\langle m_{t} \rangle - m_{0}$ for
$\pi^{+}$, K$^{+}$, p (full symbols) and their antiparticles (open
symbols) \cite{sqm04_christoph}.} \label{fig:mt-energy}
\end{minipage}
\end{figure}

An effect of the increasing baryon density is seen in the strong
decrease of the $\overline{\rm{B}}$/B ratios towards lower energies
(fig.~\ref{fig:bbar-ratios}). Although statistical models always
reproduce the ordering by strangeness content and the overall energy
dependence, a puzzling effect is hidden in this plot: The
$\overline{\Lambda}/\overline{\rm {p}}$ ratio at midrapidity is
close to/above 1 for all energies and even increases towards lower
energies (fig.~\ref{fig:alambda-ap-ratio}).
These values can not be reproduced by
statistical models. Whether an increased antiproton absorption at
lower energies might be an explanation remains an open issue since
the similar inverse slopes of the $p_{t}$-spectra measured for
protons and antiprotons (fig.~\ref{fig:mt-energy}) are not expected
in such a scenario.
\begin{figure}[htb]
\begin{minipage}[t]{75mm}
\includegraphics[width=75mm]{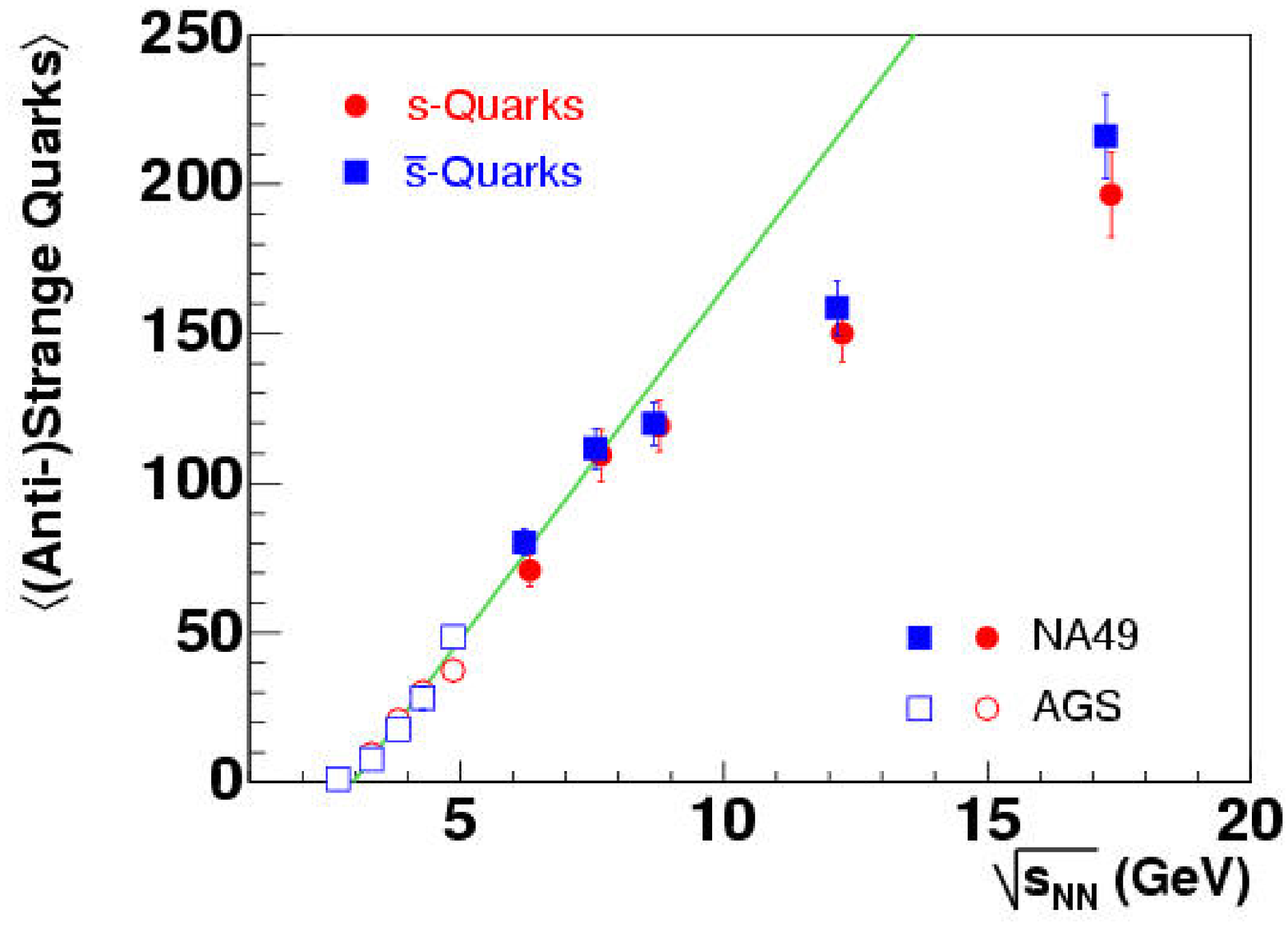}
\vspace{-0.9cm}\caption{Energy dependence of total multiplicity of
(anti-)strange constituent quarks (s-quarks: K$^{+}$,
$\overline{\rm{K^{0}}}$, $\Lambda$ including $\Sigma^{0}$, $\Xi$,
$\Omega$, and $\Sigma^{\pm}$; antiparticles for $\overline{\rm
{s}}$-quarks) \cite{sqm04_christoph}. The line gives a linear
extrapolation of the yields at low energies.}
\label{fig:total-strangeness}
\end{minipage}
\hspace{\fill}
\begin{minipage}[t]{75mm}
\includegraphics[width=75mm]{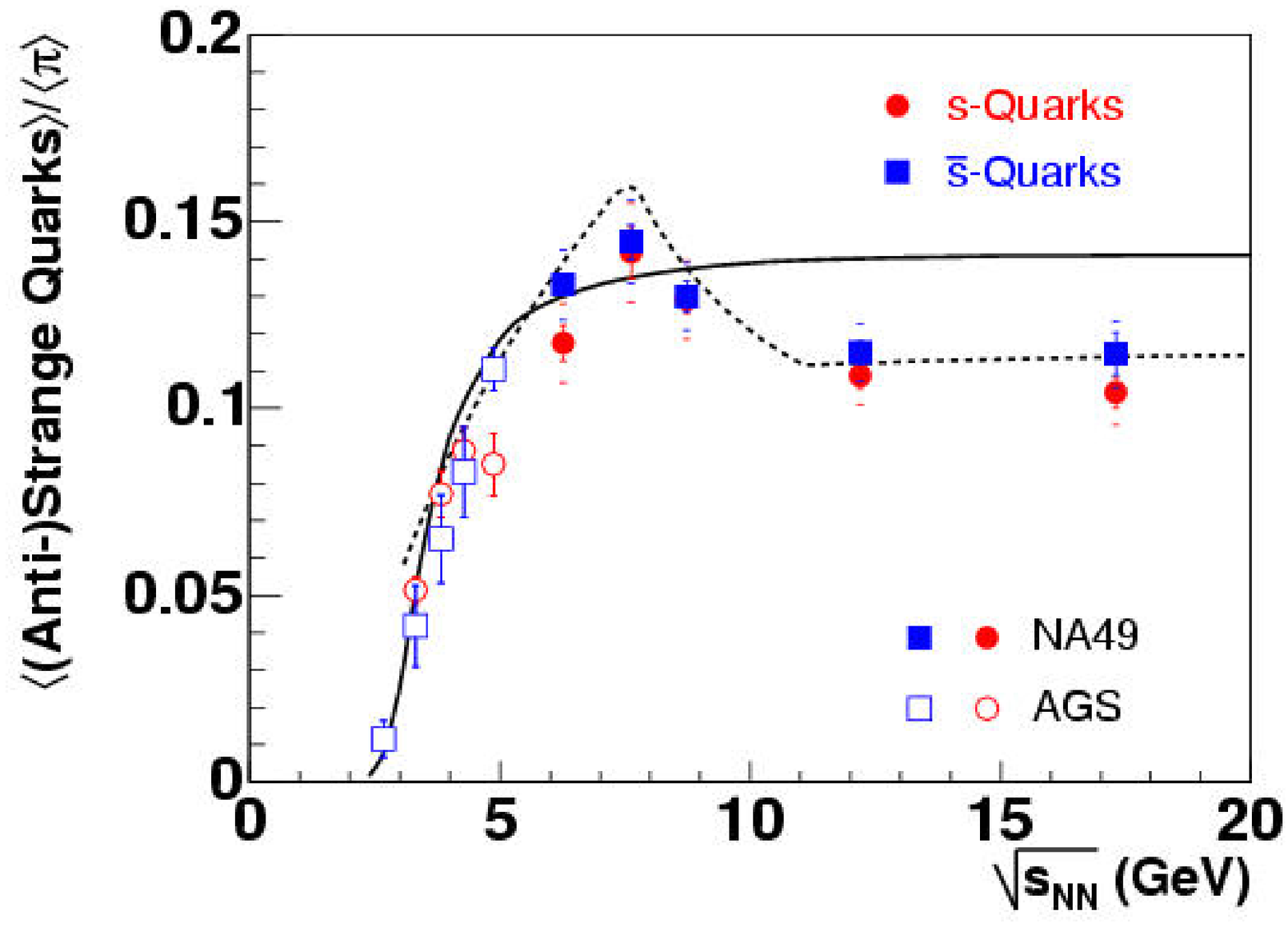}
\vspace{-0.9cm}\caption{Relative strangeness production versus
energy \cite{sqm04_christoph}. The solid line shows a statistical
hadron gas calculation with a smooth change of T and $\mu_{\rm{B}}$
and $\gamma_{s}=1$ \cite{hg-redlich}, the dashed line a prediction
assuming a first order phase transition to a partonic phase at 30$A$
GeV beam energy \cite{marek}.} \label{fig:ssbar_pi_ratio}
\end{minipage}
\end{figure}

As NA49 measures a large fraction of all strangeness carriers
\cite{na49_strangeness,marek_qm04} the energy dependence of total
strangeness production can be determined very accurately. The only
extrapolations needed are either small ($\Xi$, $\Omega$ and
antiparticles if not measured) and performed by using yields from
the hadron gas model \cite{becattini}, or they are based on the
assumption of isospin symmetry (K$^{0}$,$\overline{\rm{K^{0}}}$) and
an empirical factor of $(\Sigma^{\pm}+\Lambda)/\Lambda=1.6$ (to
estimate the yield of $\Sigma^{\pm}$ and antiparticles)
\cite{sqm04_christoph}. The sum of all (anti-)strange constituent
quarks in the produced hadrons is presented in
fig.~\ref{fig:total-strangeness}. As expected from strangeness
conservation one observes an agreement between the multiplicities of
s and $\overline{\rm {s}}$ quarks. This result represents an
important check of the internal consistency of the NA49 data.
Furthermore, fig.~\ref{fig:total-strangeness} demonstrates that the
increase of strangeness production with energy slows down at about
30$A$ GeV. Also the rate of increase of pion production grows
slightly in this region \cite{marek_qm04}. Therefore, a maximum of
the relative strangeness production is obtained around this energy
(fig.~\ref{fig:ssbar_pi_ratio}) while for higher energies extending
up to RHIC the ratio remains stable (not shown). Although the rapid
increase followed by a saturation is described by statistical hadron
gas models \cite{pbm,hg-redlich} assuming a smooth change of
temperature and baryochemical potential and full strangeness
saturation ($\gamma_{s}$=1), such a distinct maximum cannot be
explained. Also microscopic transport models such as UrQMD and HSD
fail to describe this observation \cite{urqmd-hsd}. On the other
hand, the maximum at about 30$A$ GeV was predicted as a signal of
the onset of deconfinemet \cite{marek}.

Additional support for a change in the production mechanism for
hadrons in the lower SPS energy range comes from the investigation
of $p_{t}$-spectra. Fig.~\ref{fig:mt-energy} presents the energy
dependence of mean transverse masses for pions, kaons and protons.
All show a steep rise followed by a flattening in the same range in
which the change in total and relative strangeness production is
observed. Where measured, transverse masses agree well for particles
and their antiparticles. Such a step-like behavior is not explained
by microscopic transport calculations \cite{urqmd-hsd} but is
consistent with assuming a first order phase transition with the
corresponding change of the equation of state \cite{t-kaon_hydro}.

\begin{figure}[htb]
\begin{minipage}[t]{75mm}
\includegraphics[width=75mm]{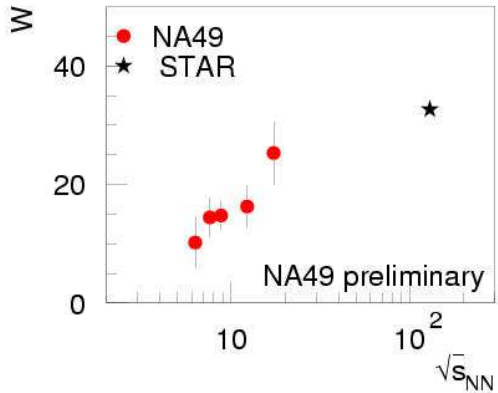}
\end{minipage}
\hspace{\fill}
\begin{minipage}[t]{75mm} \vspace{-6cm}
\caption{Energy dependence of the normalized decrease of the width
of the balance function
$W=\left(\frac{\langle\Delta\eta\rangle_{\rm{shuffling}}-\langle\Delta\eta\rangle_{\rm{data}}}
{\langle\Delta\eta\rangle_{\rm{shuffling}}}\right)\cdot 100$ for
charged particles at mid-pseudorapidity (window of 1.4 units, center
of pseudorapidity range slightly forward shifted).}
\label{fig:bf-energy}
\end{minipage}
\end{figure}

The study of particle correlations might reveal more information
about the hadronization process. Fig.~\ref{fig:bf-energy} shows new
results from an analysis of the energy dependence of the width of
the balance function of charged particles at midrapidity. In order
to compare the different energies, the width of the balance function
is normalized to shuffled events in which all correlations are
removed and for which the width has its maximum value allowed by
charge conservation. This way the acceptance dependence of the
measurement is largely removed. $W$ would increase for a decreasing
width of the balance function predicted for delayed hadronization.
Fig.~\ref{fig:bf-energy} shows that $W$ increases with energy, with
possibly a plateau in the SPS energy range. However, before
interpreting these results in terms of such a model, a more thorough
investigation of the effect of using different phase space windows
at each energy is needed.

\begin{figure}[htb]
\begin{minipage}[t]{75mm}
\includegraphics[width=60mm]{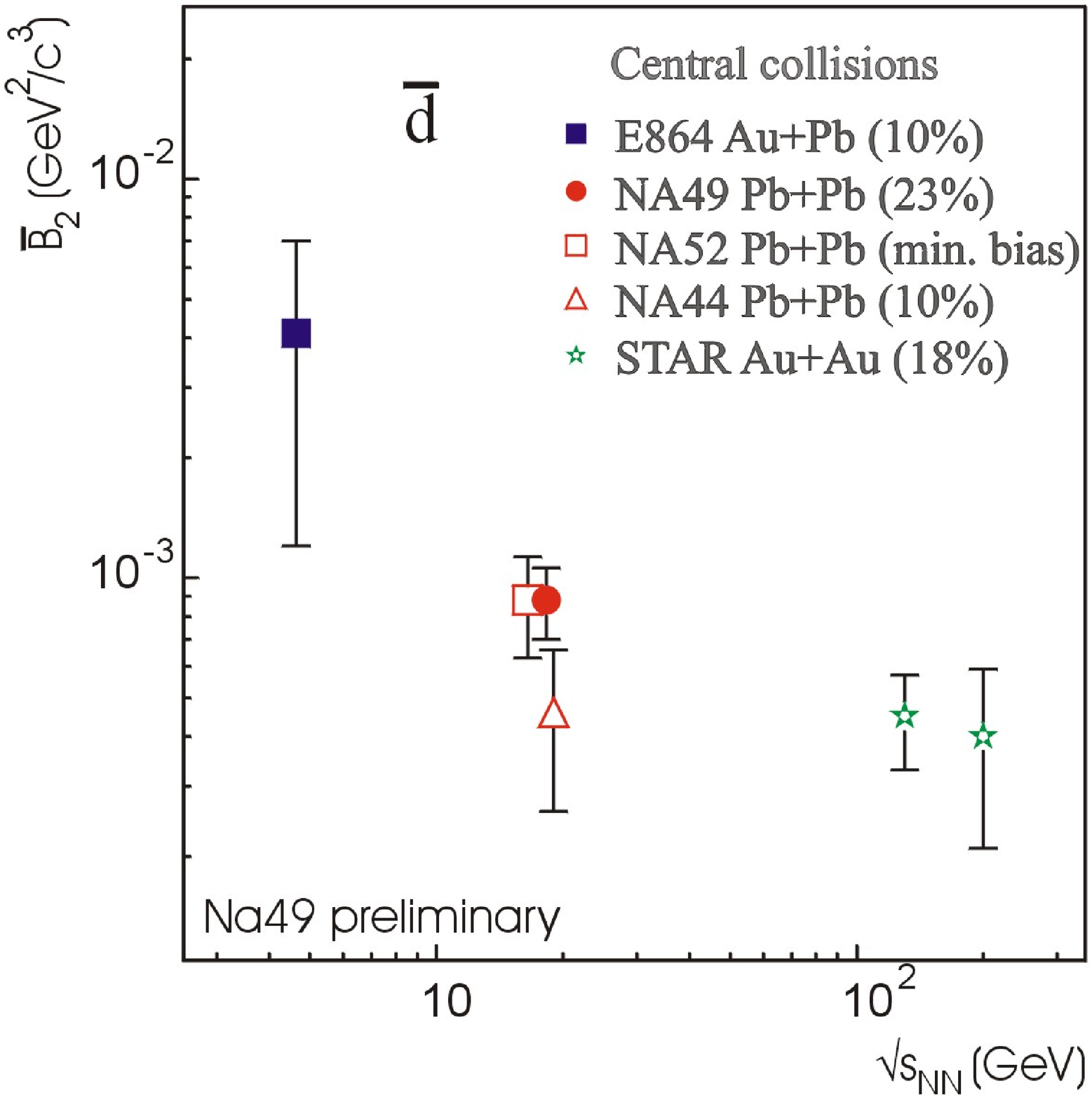}
\end{minipage}
\hspace{\fill}
\begin{minipage}[t]{75mm}
\includegraphics[width=60mm]{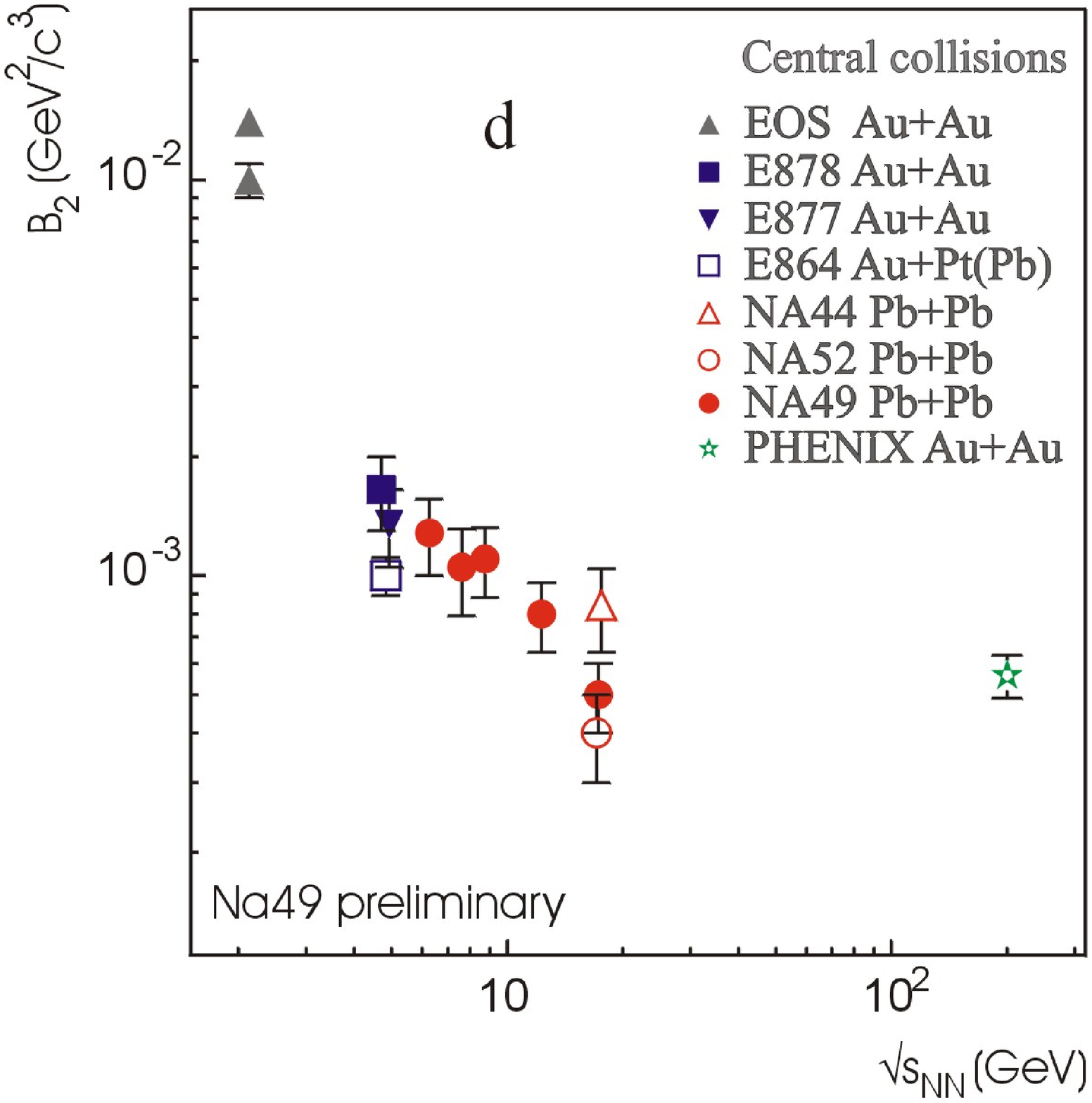}
\end{minipage}
\vspace{-0.9cm}\caption{Energy dependence of the coalescence
parameter $B_{2}$ calculated from (anti-) deuteron and (anti-)proton
spectra.} \label{fig:ppbar-ddbar}
\end{figure}

The balance function is largely dominated by mesons, whereas a
coalescence analysis of deuterons and protons can provide
information on coherence volumes for baryons. The coalescence
parameter $B_{2}$ determined from (anti-)deuteron and (anti-)proton
spectra is inversely correlated to the coherence volume
\cite{na49_deuterons}. A continuous decrease of $B_{2}$ and thus an
increase of the correlation volume is observed towards higher
energies for both, particles and antiparticles
(fig.~\ref{fig:ppbar-ddbar}).

\section{Summary and conclusion}

NA49 has systematically investigated the dependence of hadron
production on system-size/ centrality and energy, the former in
particular at the top-SPS beam energy of 158$A$ GeV, the latter
mostly for central Pb+Pb collisions. At top-SPS energies, strongly
interacting matter of high energy density is created in central
Pb+Pb collisions and the hadrochemical freeze-out occurs close to
the predicted phase boundary. For collisions of nuclei with smaller
$A$ or for peripheral Pb+Pb reactions, the results are consistent
with assuming the creation of several smaller systems and an earlier
freeze-out. For a large set of observables it was shown that the
created matter behaves in a similar manner as observed at RHIC.
However, when decreasing the beam energy distinct changes in hadron
production properties are observed around 30$A$ GeV. Although the
strongly increasing baryon density clearly influences particle
production, the observed structures can best be explained by models
assuming the onset of deconfinement at the lower SPS energies.

\vspace{0.15cm}

Acknowledgements: This work was supported by the US Department of
Energy Grant DE-FG03-97ER41020/A000, the Bundesministerium fur
Bildung und Forschung, Germany, the Virtual Institute VI-146 of
Helmholtz Gemeinschaft, Germany, the Polish State Committee for
Scientific Research (1 P03B 097 29, 1 P03B 121 29,  2 P03B 04123),
the Hungarian Scientific Research Foundation (T032648, T032293,
T043514), the Hungarian National Science Foundation, OTKA,
(F034707), the Polish-German Foundation, the Korea Science \&
Engineering Foundation (R01-2005-000-10334-0) and the Bulgarian
National Science Fund (Ph-09/05).

\end{document}